\documentclass[traditabstract]{aa}

\usepackage{amsmath}
\usepackage{graphicx}
\usepackage{amssymb}
\usepackage{natbib}
\usepackage{bm}
\usepackage{bbold}
\bibpunct{(}{)}{;}{a}{}{,}

\newcommand{\mubold}{\mbox{\boldmath$\mu$}}
\newcommand{\thetabold}{\mbox{\boldmath$\theta$}}
\newcommand{\Sigmabold}{\mbox{\boldmath$\Sigma$}}

\newcommand{\alphabold}{\mbox{\boldmath$\alpha$}}
\newcommand{\Phibold}{\mbox{\boldmath$\Phi$}}

\begin{document}

\title{Signal detection for spectroscopy and polarimetry}
\author{A. Asensio Ramos \& R. Manso Sainz}

\offprints{aasensio@iac.es}

\institute{Instituto de Astrof\'{\i}sica de Canarias, 38205, La Laguna, Tenerife, Spain; \email{aasensio@iac.es}
\and
Departamento de Astrof\'{\i}sica, Universidad de La Laguna, E-38205 La Laguna, Tenerife, Spain
}
\date{Received ; Accepted}
\titlerunning{Signal detection for spectroscopy and polarimetry}
\authorrunning{Asensio Ramos \& Manso Sainz}
% 
% 
% % abstract cannot exceed 300 words
\abstract{The analysis of high spectral resolution spectroscopic and spectropolarimetric observations constitute a very
powerful way of inferring the dynamical, thermodynamical, and magnetic properties of distant objects.
However, these techniques are photon-starving, making it difficult to use them for all purposes. One
of the problems commonly found is just detecting the presence of a signal that is buried on the noise
at the wavelength of some interesting spectral feature. This is specially relevant for spectropolarimetric
observations because typically, only a small fraction of the received light is polarized. We present in this note
a Bayesian technique for the detection of spectropolarimetric signals. The technique is based on the application
of the non-parametric relevance vector machine to the observations, which allows us to compute the
evidence for the presence of the signal and compute the more probable signal. 
The method would be suited for analyzing data from experimental instruments onboard space missions 
and rockets aiming at detecting spectropolarimetric signals in unexplored regions of the spectrum such as the 
Chromospheric Lyman-Alpha Spectro-Polarimeter (CLASP) sounding rocket experiment.}


\maketitle

%%%%%%%%%%%%%%%%%%%%%%%%%%%%%%%%%%%%%%%%
%%%%%%%%%%%%%%%%%%%%%%%%%%%%%%%%%%%%%%%%
% INTRODUCTION
%%%%%%%%%%%%%%%%%%%%%%%%%%%%%%%%%%%%%%%%
%%%%%%%%%%%%%%%%%%%%%%%%%%%%%%%%%%%%%%%%
\section{Introduction}
Spectroscopy and spectropolarimetry are two of the most important techniques in
the observational astrophysics toolbox. 
By recording the intensity and polarization state of light at each wavelength we
get a complete\footnote{Or nearly so: see \cite{harwit03} and \cite{uribe11}.}
characterization of the state of the light from the observed object, and from
its analysis we may infer all the available information on the chemical,
thermodynamical, and magnetic properties of the plasma that emitted that light.
In some cases, even the mere detection of a given spectral or polarimetric
feature may provide fundamental constraints on the observed object.
For example, just the measurement of a linear polarization signal from an
unresolved object may imply strong constraints on its geometry (it cannot be
spherically symmetric), the presence of an organized magnetic field, or both.

The main drawback of spectroscopy and spectropolarimetry is that they are often
photon-starving techniques.
Spectroscopic observations are characterized by the spectral resolution of the
spectrograph $R=\lambda/\Delta\lambda$ ($\Delta\lambda$ is the wavelength
interval within a resolution element observed at the wavelength $\lambda$)
which,
in the optical and infrared, may typically range $R\sim 1000-1000000$ (for
low-resolution night-time spectrographs or solar spectrographs, respectively).
On the other hand, the fraction of polarized photons $P$ in a light beam is
$P\sim 1$-$10\%$ for strongly polarized sources and, typically, $P\lesssim
10^{-3}$. 
Even worse, polarization is subject to cancellations and $P$ decreases rapidly
for low resolution observations.
As a consequence, even with the largest telescopes and the most efficient
instrumentation the number of (polarized) photons finally reaching a resolution
element of the detector may be very low and close to the noise levels (either
the photon noise or the noise of the detection devices), rendering the detection
of the signal difficult. 
In those cases, the presence of a spectral pattern is often determined from
heuristic or somehow subjective arguments.
Tipically, some kind of filtering is applied to the data to enhance the possible
signal, which is then identified graphically, by simple visual inspection or by
fitting of an appropriate parametric function. 
A quantitative assessment of the quality of the detection or an objective
estimation confidence intervals is then lacking or impossible.

%The mere detection of a spectroscopic and/or spectropolarimetric signal in a certain spectral
%region can give extense information about the physical properties inside astrophysical
%plasmas. Sometimes, the limited size of our telescopes, the presence of noise (either
%the photon noise or the noise present in the detection devices) and the
%inherently weak signals that one is interested in in extreme cases, renders the detection of a signal 
%a very difficult task. 

%Unfortunately, the detection of a signature is often based on subjective arguments.
%Typically, the signal is pushed out from its noise burial using any of the filtering methods 
%available in the literature. Once at this point, the signal of interest is isolated
%by eye graphically or it is extracted fitting it with an appropriate parametric
%function. There is not a quantitative measure of the quality of the signal detection or an
%estimation of the confidence intervals in both methods.

In this paper we apply a Bayesian non-parametric regression method for the extraction of
spectroscopic and/or spectropolarimetric signals (or any other one-dimensional signal) from
noisy observations. The method is based on relevance vector machines \cite[RVM;][]{tipping00}, a 
Bayesian version of the support vector machine machine learning technique. Several
fundamental advantages are gained. First, we are able to quantify signal detection by
computing the evidence ratio between two models: one that contains the signal of interest
plus noise and one in which there is only noise. Second, the complexity of the signal
is automatically adapted to the information present in the observations. Observations with low noise
will facilitate the inference of minute details in the signal of interest, while
very noisy observations will favor simpler (and typically smoother) signals. Finally, we obtain
an estimation of the signal, together with error bars.
We demonstrate the formalism with its application to 
synthetic and real data.

\section{General considerations}
Consider the detection of a spectroscopic signal $I(\lambda)$ (equivalently for spectropolarimetric
signals) in an observation perturbed with Gaussian noise with zero mean and
variance $\sigma^2$. In principle, two possibilities may be contemplated. One, what we term model ${\cal M}_1$,
that there is indeed a signal on the observations $I(\lambda)$ and that it is corrupted with
Gaussian noise; the other, model ${\cal M}_0$, that there is not such a signal at all, only Gaussian noise.
The two options give the following models for the observed signal:
\begin{subequations}
\begin{eqnarray}
d(\lambda_i) &=& I(\lambda_i) + \epsilon_i, \\
d(\lambda_i) &=& \epsilon_i.
\end{eqnarray}
\end{subequations}
where we make explicit that the observed signal is sampled at a set of wavelength 
points $\{\lambda_i\}_{i=1}^N$.

If a good parametric model depending on the vector of parameters $\thetabold$ is available
for the expected signal $I(\lambda;\thetabold)$, the most straightforward way to proceed in order to test for the 
presence of the signal on given observation (that we represent by the vector $\mathbf{d}$, built by stacking
the observed fluxes at all observed wavelength points) is to compute likelihood
ratio \citep{cox06}:
\begin{equation}
R = \frac{p(\mathbf{d}|\thetabold_\mathrm{ML},\mathcal{M}_1)}{p(\mathbf{d}|\mathcal{M}_0)},
\end{equation}
where the likelihood for the model $\mathcal{M}_1$ is given by
\begin{equation}
p(\mathbf{d}|\thetabold,\mathcal{M}_1) = \prod_{i=1}^N (2\pi\sigma^2)^{-1/2} 
\exp \left[- \frac{\left[d(\lambda_i) - I(\lambda_i;\thetabold) \right]^2}{2\sigma^2} \right],
\label{eq:likelihood}
\end{equation}
and it is evaluated at the parameters that maximize it.
Note that the likelihood is the product of $N$ Gaussians because of the noise
model we have chosen (uncorrelated noise with zero mean and variance $\sigma^2$).
Likewise, the likelihood for the model $\mathcal{M}_0$ is 
\begin{equation}
p(\mathbf{d}|\mathcal{M}_0) = \prod_{i=1}^N (2\pi\sigma^2)^{-1/2} \exp \left[- \frac{d(\lambda_i)^2}{2\sigma^2} \right].
\label{eq:likelihood_nosignal}
\end{equation}
The decision about the presence of the signal is done in terms of the ratio
at different confidence levels \citep[see][]{cox06} and the signal that
obtained with parameters $\thetabold_\mathrm{ML}$ is the maximum likelihood 
signal.

In spite of the simplicity, there is a fundamental problem in the likelihood ratio. Using
the maximum likelihood value of the parameters, one is not taking into account the uncertainty 
about $\thetabold$. One of the consequences is that it is possible to promote complicated models
if the number of parameters is sufficiently large, leading to overfitting. In other words,
in complex models, we can fit the noise so that signal is always detected.
That is the reason why model comparison (and, consequently, signal detection) is done
in the Bayesian formalism through the evidence ratio (or Bayes ratio)
\citep[e.g.,][]{jeffreys61,kass_raftery95,gregory05,trotta08,asensioramos_modcomp12}:
\begin{equation}
R = \frac{p(\mathbf{d}|\mathcal{M}_1)}{p(\mathbf{d}|\mathcal{M}_0)},
\label{eq:evidence_ratio}
\end{equation}
which gives the ratio of the probability that the observed data has been generated
by a model with a signal and the probability that the observed data is just compatible with
pure noise. These ratios can be transformed into strengths of belief using
the modified Jeffreys scale that has been presented by \cite{jeffreys61}, \cite{kass_raftery95} or \cite{gordon_trotta07}.

Two main differences appear between the evidence ratio and the likelihood ratio. The first one
is that model comparison is done with the evidences, in which parameters have been integrated:
\begin{equation}
p(\mathbf{d}|\mathcal{M}_1) = \int d\thetabold \, p(\mathbf{d}|\thetabold,\mathcal{M}_1) p(\thetabold|\mathcal{M}_1),
\label{eq:evidence_theta}
\end{equation}
so uncertainties in $\thetabold$ are taken into account. The second one is the standard inclusion
of a prior distribution for the parameters, which works as a regularizing term.

% From a Bayesian point of view, the Bayes theorem describes how the a-priori state of knowledge about
% the model parameters is updated once data is incorporated into the problem:
% \begin{equation}
% p(\thetabold|\mathbf{d},\mathcal{M}_i) = \frac{p(\mathbf{d}|\thetabold,\mathcal{M}_i) p(\thetabold|\mathcal{M}_i)}{p(\mathbf{d}|\mathcal{M}_i)},
% \end{equation}
% where $p(\thetabold|\mathbf{d},\mathcal{M}_i)$ is the posterior distribution for the model
% parameters (what one knows about the parameters after presenting the data in the
% inference process), $p(\mathbf{d}|\thetabold,\mathcal{M}_i)$ is the likelihood (the information
% about the parameters encoded on the observations), $p(\thetabold|\mathcal{M}_i)$ 
% is the prior distribution for the model parameters (the original knowledge about the
% model parameters). The term $p(\mathbf{d}|\mathcal{M}_i)$ is the 
% evidence or marginal posterior, whose role is to keep the posterior distribution normalized. Therefore:
% 
% Particularizing to our case, model comparison (signal detection in our
% framework) is then performed by computing the ratio \citep{jeffreys61}:

\section{Bayesian signal detection with non-parametric models}
Parametric models are appropriate when one is confident about the shape of the expected
signal. For instance, it can be used to detect a spectral 
line that is known to have Gaussian
shape although the precise position, broadening and amplitude are unknown. However, this
is not often the case, at least for the following two reasons. First, many of the interesting cases are those
in which the observed signal cannot be reproduced with our models, constituting a potential 
source of new phenomena (e.g., several velocity components in the spectral line generate a very
complex pattern that is difficult to anticipate). Second, it might be that an 
observations is made with the aim of detecting a signal that has never been 
observed, making it difficult to propose a parametric model
that can explain its exact shape. 

To overcome the potential failure of parametric models, non-parametric regression models have also 
been developed in the recent years. Non-parametric regression relies on the application of a 
sufficiently general function that depends only on observed quantities and that is used to 
approximate the observations. The signal detection scheme we have developed is based on the application of the relevance vector 
machine \cite[RVM;][]{tipping00}, a Bayesian update of the support vector machine machine 
learning technique \citep{vapnik95}. In this case, the general function is
just a linear combination of kernels:
\begin{equation}
I(\lambda;\mathbf{w}) = \sum_{j=1}^M w_j K_j(\lambda),
\label{eq:model}
\end{equation}
where the $K_j(\lambda)$ functions are arbitrary and defined in advance and $w_j$ is the
weight associated to the $j$-th kernel function. This functional form 
is also known as a linear regressor. The parameters we infer from the data appear
linearly in the model once the kernel functions are fixed. For instance, if the kernel functions
are chosen to be polynomials, one ends up with a standard polynomial regression.
The main advantage of non-parametric regression is that the model automatically adapts to
the observations. For this adaption to occur, the basis functions should ideally
capture part of the behavior of the signal. Together with the fact that the number of basis 
functions that one can include into the linear regression can be arbitrarily large (even potentially infinite, in some cases),
this constitutes a very powerful model for any unknown signal.

\subsection{Hierarchical modeling}
The linear regression problem is usually solved by computing the value of the weights $w_j$ that minimize
the $\ell_2$-norm between the observations and the predictions \citep[e.g.,][]{numerical_recipes86}. 
In other words, the value of
the $w_j$ are the solution to the least-squares problem. However, it is known that the
least-squares solution leads to severe overfitting and renders the method useless. \cite{tipping00}
considered to overcome the overfitting by pursuing a hierarchical Bayesian solution to the linear
regression problem. The aim is to use the available data to compute the posterior distribution function 
for the vector of weights $\mathbf{w}$ and the noise variance $\sigma^2$ (that will be
estimated from the same data). Therefore, a direct application of the Bayes theorem will give:
\begin{equation}
p(\mathbf{w},\sigma^2|\mathbf{d},\mathcal{M}_1) = \frac{p(\mathbf{d}|\mathbf{w},\sigma^2,\mathcal{M}_1) p(\mathbf{w},\sigma^2|\mathcal{M}_1)}{p(\mathbf{d|\mathcal{M}_1})},
\end{equation}
where $p(\mathbf{d}|\mathbf{w},\sigma^2,\mathcal{M}_1)$ is the likelihood function given
by Eq. (\ref{eq:likelihood}), $p(\mathbf{w},\sigma^2|\mathcal{M}_1)$ is the
prior distribution for the parameters that we define now and 
$p(\mathbf{d}|\mathcal{M}_1)$ is the evidence which, like Eq. (\ref{eq:evidence_theta}),
is given by the integral over $\mathbf{w}$ and $\sigma^2$ of the numerator of the right hand side. 
In order to simplify the notation, we drop the conditioning on $\mathcal{M}_1$ from now on because
we are focusing on the model that assumes the presence of signal.
Putting flat priors on $\mathbf{w}$ and $\sigma^2$ (i.e., $p(\mathbf{w},\sigma^2) \propto 1$) is equivalent
to the maximum-likelihood solution, which might lead to overfitting.
In order to overcome this problem, \cite{tipping00} used a hierarchical approach in which the prior
for $\mathbf{w}$ is made to depend on a set of hyperparameters $\alphabold$, which are learnt from the
data during the inference process. The final posterior distribution is then, after following the standard
procedure in Bayesian statistics of including a prior for the newly defined random variables, given by:
\begin{equation}
p(\mathbf{w},\alphabold,\sigma^2|\mathbf{d}) = \frac{p(\mathbf{d}|\mathbf{w},\sigma^2) p(\mathbf{w},\alphabold,\sigma^2)}
{p(\mathbf{d})}.
\label{eq:bayes1}
\end{equation}
Note that the likelihood does depend directly on $\mathbf{w}$ and not on the election of $\alphabold$.
Assuming that the prior for $\alphabold$ and $\sigma^2$ are independent and that the prior
for $\mathbf{w}$ depend on the hyperparameters $\alphabold$, the previous equation can be trivially modified to read:
\begin{equation}
p(\mathbf{w},\alphabold,\sigma^2|\mathbf{d}) = \frac{p(\mathbf{d}|\mathbf{w},\sigma^2) p(\mathbf{w}|\alphabold)p(\alphabold)p(\sigma^2)}{p(\mathbf{d})}.
\label{eq:bayes2}
\end{equation}
The value of the evidence, or marginal posterior, is computed to ensure that the posterior is
normalized to unit hyperarea:
\begin{equation}
p(\mathbf{d}|\mathcal{M}_1) = \int \mathrm{d}\mathbf{w} 
\mathrm{d}\alphabold  \mathrm{d}\sigma^2 \, p(\mathbf{d}|\mathbf{w},\sigma^2) p(\mathbf{w}|\alphabold) p(\alphabold) p(\sigma^2) ,
\label{eq:marginal}
\end{equation}
where the priors $p(\mathbf{w}|\alphabold)$, $p(\alphabold)$ and $p(\sigma^2)$ are still left undefined and
we have made explicit again the conditioning on $\mathcal{M}_1$ for clarity.

\subsection{Sparsity prior}
One of the fundamental ideas of RVMs is to regularize the regression problem by favoring 
the sparsest solutions, i.e., those that contain the least number of non-zero elements in $\mathbf{w}$. 
For this reason, and to keep
the analytical tractability, \cite{tipping00} decided to use a product of Gaussian functions 
for $p(\mathbf{w}|\alphabold)$:
\begin{equation}
p(\mathbf{w}|\alphabold) = \prod_{i=1}^M \mathcal{N}(w_i|0,\alpha_i^{-1}),
\end{equation}
where $\mathcal{N}(w|\mu,\sigma^2)$ is a Gaussian distribution on the variable $w$ with mean $\mu$
and variance $\sigma^2$.
Although not obvious, this prior favors small values of $\mathbf{w}$ when selecting an
appropriate prior for $\alphabold$. The reason is that, in the hierarchical scheme, the 
final prior over $\mathbf{w}$ is given by the marginalization:
\begin{equation}
p(\mathbf{w}) = \int \mathrm{d} \alphabold \, p(\mathbf{w}|\alphabold) p(\alphabold).
\end{equation}
If a Jeffreys prior is used for each $\alpha_i$ so that $p(\alpha_i) = \alpha_i^{-1}$, we end up 
with $p(w_i) \propto |w_i|^{-1}$, which clearly favors small values of $w_i$.
In essence, the form of $p(\mathbf{w}|\alphabold)$ is such that, in the limiting case 
that $\alpha_i$ tends to infinity, the marginal prior for $w_i$ is so peaked at zero that is compatible with a
Dirac delta. This means that this specific $w_i$ does not contribute to the model of Eq. (\ref{eq:model})
and can be dropped from the model without impact. This regularization proposed by \cite{tipping00}
leads to a sparse $\mathbf{w}$ vector, so an automatic relevance determination is
implemented in the method. 

% Therefore, if we
% assume that the noise variance $\sigma^2$ and the hyperparameters $\alphabold$ are known
% in advance, the Bayesian posterior results in:
% \begin{equation}
% p(\mathbf{w}|\mathbf{d},\alphabold,\sigma^2) = \frac{p(\mathbf{d}|\mathbf{w},\sigma^2) p(\mathbf{w}|\alphabold)}{p(\mathbf{d})}.
% \end{equation}

\begin{figure*}[!t]
\centering
\includegraphics[width=13.5cm,angle=-90]{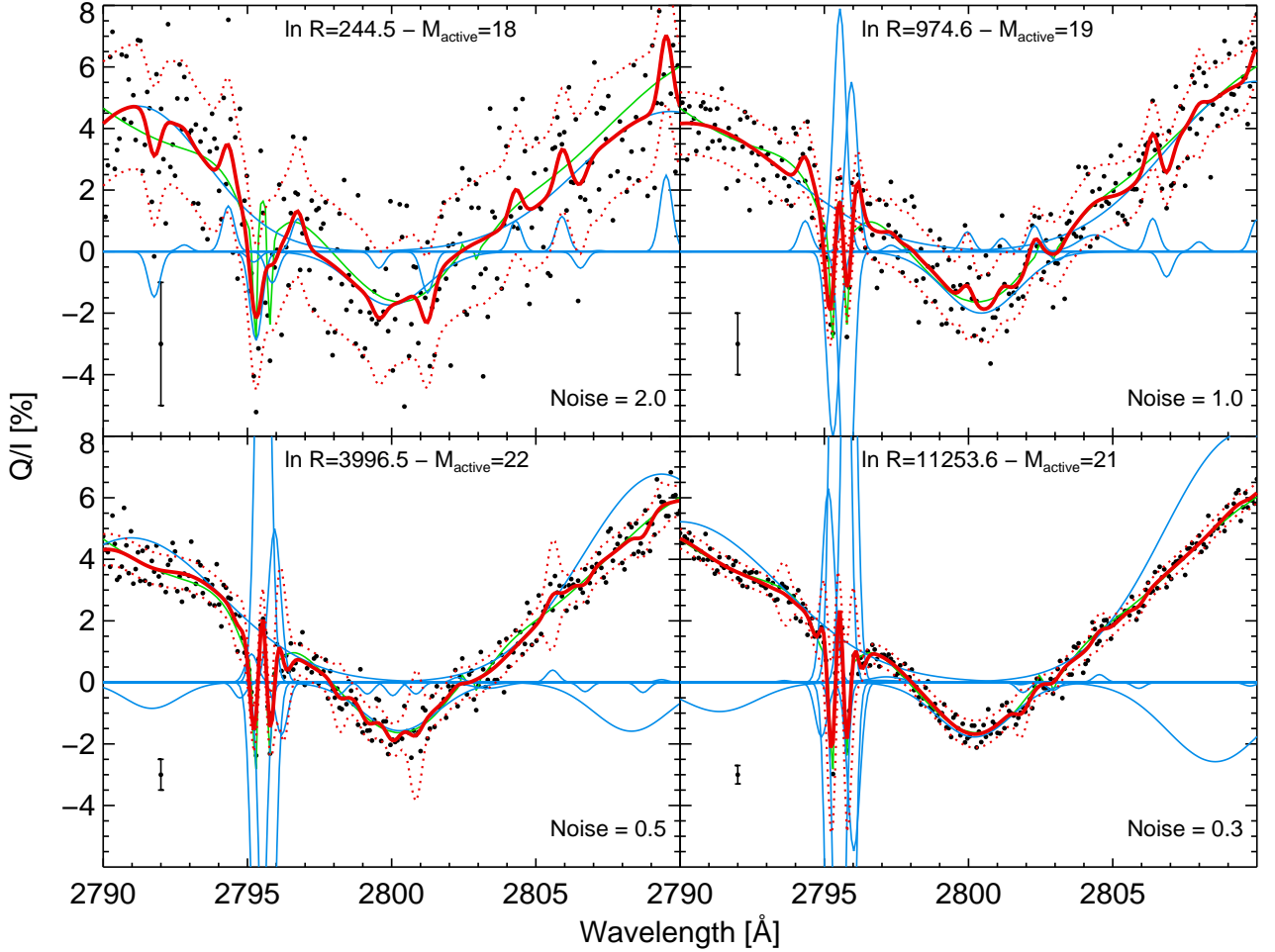}
\caption{Example of the Bayesian signal detection scheme applied to a synthetic
spectrum of the scattering polarization pattern across the Mg \textsc{ii} h and k lines obtained by \cite{belluzzi_trujillo_mg12}. The dots display the
observations with increasingly higher Gaussian noise. In order to avoid cluttering, we only
show one error bar. The solid (dotted) red curve is the mean (standard deviation) of the Gaussian predictive distribution and
is computed using Eq. (\ref{eq:predictive}). The blue curves display the
contribution of each individual kernel function, while the green curve is the original
synthetic profile. The logarithm of the evidence ratio given in Eq. (\ref{eq:evidence_ratio})
is shown for each case. Additionally, we also display $M_\mathrm{active}$, the number of active basis 
functions considered by the Relevance Vector Machine algorithm.\label{fig:mgii_belluzzi}}
\end{figure*}

\subsection{Type-II maximum likelihood}
The computation of the evidence of Eq. (\ref{eq:marginal}) is intractable. 
Looking for an analytical solution, \cite{tipping00} proceeded with a Type-II Maximum
likelihood approximation \citep[also known as empirical Bayes, generalized maximum
likelihood or evidence approximation;][]{mackay99}. The idea is that, if the posterior for the hyperparameters
$\alphabold$ and the noise variance $\sigma^2$ is fairly peaked, one can substitute their values by
their modes and simplify the expressions. Therefore, if we make the substitutions
$p(\alphabold)=\delta(\alphabold-\alphabold_\mathrm{MP})$ and $p(\sigma^2)=\delta(\sigma^2-\sigma^2_\mathrm{MP})$,
where the subindex ``MP'' refers to the maximum a-posterior values, the evidence in Eq. (\ref{eq:marginal})
simplifies to:
% Using standard probability calculus rules, the full 
% posterior in Eq. (\ref{eq:bayes2}) can be factorized as:
% \begin{equation}
% p(\mathbf{w},\alphabold,\sigma^2|\mathbf{d}) = p(\mathbf{w}|\alphabold,\sigma^2,\mathbf{d}) p(\alphabold,\sigma^2|\mathbf{d}),
% \label{eq:posterior_factorized}
% \end{equation}
% where the second term can be written, applying the Bayes theorem, as:
% \begin{equation}
% p(\alphabold,\sigma^2|\mathbf{d}) = \frac{p(\mathbf{d}|\alphabold,\sigma^2) p(\alphabold) p(\sigma^2)}{p(\mathbf{d})}.
% \label{eq:posterior_alpha_sigma}
% \end{equation}
% Note that, if we choose flat priors for $\alphabold$ and $\sigma^2$, the values $\alphabold_\mathrm{MP}$ and $\sigma^2_\mathrm{MP}$
% that maximize $p(\alphabold,\sigma^2|\mathbf{d})$ are the same ones that maximize the marginal likelihood
% $p(\mathbf{d}|\alphabold,\sigma^2)$, which is obtained by computing the integral of Eq. (\ref{eq:marginal})
% using the Dirac delta priors $p(\alpha_i)=\delta(\alpha_i-\alpha_\mathrm{MP})$:
\begin{equation}
p(\mathbf{d}|\mathcal{M}_1) = \int \mathrm{d}\mathbf{w} \, p(\mathbf{d}|\mathbf{w},\sigma^2_\mathrm{MP}) p(\mathbf{w}|\alphabold_\mathrm{MP}),
\label{eq:evidence_final}
\end{equation}
which is now Gaussian with zero mean and covariance matrix:
\begin{equation}
\Sigmabold_\mathrm{ev} = \left( \sigma_\mathrm{MP}^2 \mathbb{1} + \Phibold \mathbf{A}^{-1} \Phibold^T \right)^{-1}
\end{equation}
where $\mathbf{A}=\mathrm{diag}(\alpha_1,\alpha_2,\ldots,\alpha_M)$ is a diagonal matrix with the $\alphabold_\mathrm{MP}$
vector in the main diagonal, $\mathbb{1}$ is the identity matrix and $\Phibold$ is the $N \times M$ matrix with elements $\Phibold_{ij}=K_j(\lambda_i)$.
The strategy to follow is then to compute the value of the elements of $\alphabold_\mathrm{MP}$ (and $\sigma^2_\mathrm{MP}$
if one also wants the noise variance estimated from the data) that maximize the evidence given by 
Eq. (\ref{eq:evidence_final}) and fix them to the inferred values to proceed. The evidence is used
afterwards for model comparison purposes.

\subsection{Predictive distribution}
Given the information gained from the data about $\alphabold$, $\sigma^2$ and $\mathbf{w}$, the
predicted value $I_\star$ at an arbitrary wavelength $\lambda_\star$ is a random variable. Its
distribution, known as predictive distribution, is given by \citep[e.g.,][]{gregory05}:
\begin{equation}
p(I_\star|\mathbf{d}) = \int \mathrm{d} \alphabold \mathrm{d} \mathbf{w} \mathrm{d} \sigma^2 \, p(I_\star|\mathbf{w},\sigma^2) 
p(\mathbf{w},\alphabold,\sigma^2|\mathbf{d}),
\end{equation}
which is just the integral of the likelihood for a new value $I_\star$ associated with $\lambda_\star$
weigthed by the posterior distribution for all the parameters. 
Under the Type-II maximum likelihood approach that we have applied before, the integral
over $\alphabold$ and $\sigma^2$ can be carried out analytically so:
\begin{equation}
p(I_\star|\mathbf{d}) = \int \mathrm{d} \mathbf{w} \, p(I_\star|\mathbf{w},\sigma_\mathrm{MP}^2) p(\mathbf{w},\alphabold_\mathrm{MP},\sigma_\mathrm{MP}^2|\mathbf{d}).
\label{eq:predictive_distribution}
\end{equation}
The result of the integral turns out to be a Gaussian distribution with the following mean and variance:
\begin{eqnarray}
\mu_\star &=& I(\lambda_\star;\mubold) \nonumber \\
\sigma_\star^2 &=& \sigma_\mathrm{MP}^2 + \mathbf{f}^T \Sigmabold \mathbf{f},
\label{eq:predictive}
\end{eqnarray}
where $\mathbf{f}=[K(\lambda_\star-\lambda_1),\ldots,K(\lambda_\star-\lambda_N)]^T$ and
\begin{eqnarray}
\mubold &=& \left( \Phibold^T \Phibold + \sigma^2_\mathrm{MP} \mathbf{A} \right)^{-1} \Phibold^T \mathbf{d} \nonumber \\
\Sigmabold &=& \sigma^2_\mathrm{MP} \left( \Phibold^T \Phibold + \sigma^2_\mathrm{MP} \mathbf{A} \right)^{-1},
\label{eq:mu_sigma}
\end{eqnarray}
with the $\mathbf{A}$ and $\Phibold$ matrices defined above.

\begin{figure}[!t]
\includegraphics[width=\columnwidth]{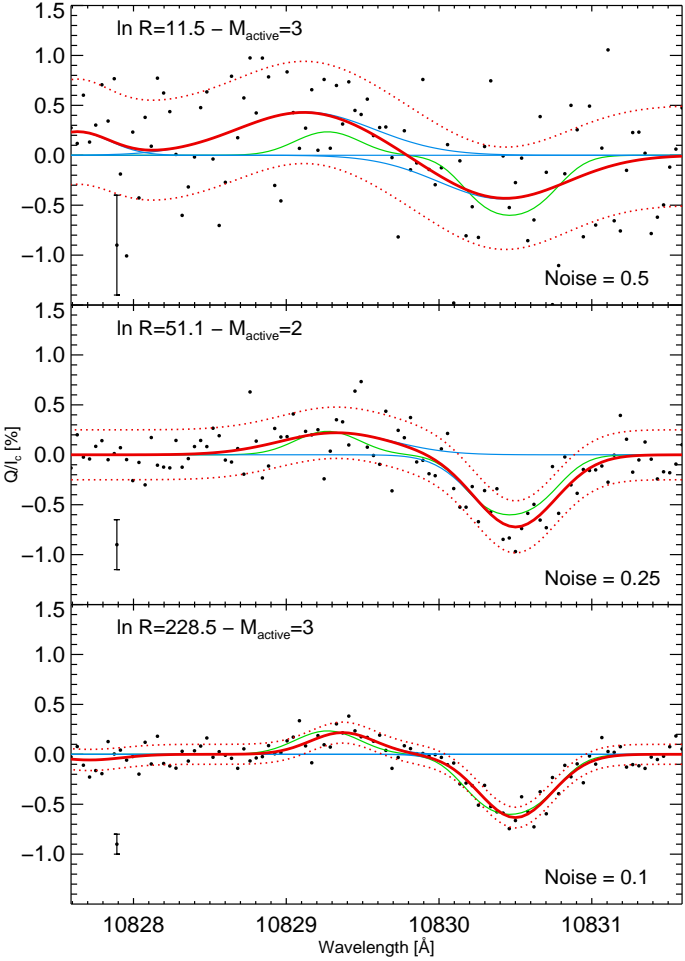}
\caption{Example of the Bayesian signal detection scheme applied to a synthetic
spectrum of the He~{\sc i} 10830 \AA\ multiplet obtained with Hazel \citep{asensio_trujillo_hazel08}. The dots display the
observations with increasingly higher Gaussian noise. In order to avoid cluttering, we only
show one error bar. The solid red curve is the mean of the predictive distribution, 
together with the error bars shown in red dotted lines. The blue curves display the
contribution of each individual kernel function, while the green curve is the original
profile.\label{fig:hei_theory}}
\end{figure}

% \subsection{Model without signal}
% Signal detection is based on comparing the evidence for a model with signal and a
% model without signal. Making the assumption that there is no signal on the observations is equivalent to 
% assuming that $p(\mathbf{w}|\alphabold)=\delta(\mathbf{w})$. Since the prior for $\mathbf{w}$ does
% not depend on the hyperparameters $\alphabold$, the evidence given by Eq. (\ref{eq:marginal}) simplifies to:
% \begin{equation}
% p(\mathbf{d}) = \prod_{n=1}^N (2\pi\sigma^2)^{-1/2} \exp \left[- \frac{y_n^2}{2\sigma^2} \right],
% \label{eq:likelihood_nosignal}
% \end{equation}
% with the noise variance estimated just as the sample variance in the observed data.

\subsection{Summary}
Summarizing, one computes the values of $\alphabold_\mathrm{MP}$ and $\sigma^2_\mathrm{MP}$ that
maximize the evidence of Eq. (\ref{eq:evidence_final}) and uses these values to estimate
the mean and variance of the predicted value at an arbitrary new point $\lambda_\star$
using Eqs. (\ref{eq:predictive}). During the optimization of the evidence, the RVM
algorithm devised by \cite{tipping00} discards all the functions contributing to the regression function of Eq. (\ref{eq:model})
whose value of $\alpha_i$ becomes very large. If $\alpha_i$ becomes very large, this means that the 
kernel function associated to $w_i$ is not needed (thus the method automatically selected which basis 
functions are needed depending on the noise level)\footnote{A signal detection code based on the routines of \cite{tipping00} can
be freely downloaded from http://www.iac.es/project/magnetism/signal\_detection.}.

\section{Applications}
We present the characteristics of the method with applications to several signal
detection examples. We start with some synthetic cases to verify the robustness
of the method to different noise levels. Then, we apply it to a few realistic
cases. Although the RVM method is able to infer the noise variance $\sigma^2$
from the data, we prefer in this paper to give this as an input by setting $\sigma_\mathrm{MP}^2=\sigma^2$
in order to show the ability of the method to extract the signal when the noise
level is correctly estimated. In any case, we have tested in all cases that, if the value of
$\sigma_\mathrm{MP}^2$ is inferred from the maximization of Eq. (\ref{eq:evidence_final}), its value
is quite similar to the original noise variance introduced in the experiments.

\subsection{Synthetic data}

\subsubsection{Linear polarization of the Mg \textsc{ii} h and k lines}
The linear polarization signal in the Mg \textsc{ii} h and k lines around 2800 \AA\ produced
by coherent scattering is expected to be large given the large anisotropy of the ultraviolet (UV) 
radiation field in this spectral region. However, the observation of this UV window cannot
be accomplished from the ground and one has to use space-borne observatories. Consequently,
it is expected that the detection of such signals in the future will be a technical challenge.

In order to test our signal detection procedure, we have used
the theoretical results of \cite{belluzzi_trujillo_mg12} as a testbench. They synthesize the emergent $Q/I$ across the
h and k lines taking into account partial redistribution (PRD) and $J$-state interference effects 
in the FAL-C semiempirical atmosphere of \cite{fontenla_falc93} for an observation at $\mu=0.1$,
with $\mu$ the heliocentric angle. The synthetic curve is shown as a green curve in Fig. \ref{fig:mgii_belluzzi}.
The calculations are done in an adaptive wavelength axis so that the sampling close to the line cores is
finer than away from them. Since this will not be the case in real observations,
we resample the profile at fixed intervals of 80 m\AA\ and add different noise
levels characterized by their standard deviation, quoted in the lower right
corner of each panel. These figures are representative of instruments like
IRIS \citep{depontieu_iris09}, which will observe these very same lines but without polarimetric
capabilities.

The previous formalism is applied using a basis set 
consisting of Gaussian functions centered at each observed point and with
widths ranging from 0.3 \AA\ to 10 \AA\ in 20 steps of 0.5 \AA\ plus a constant
function to allow for a continuum bias. The reason to allow for such a variety
of basis functions is to simultaneously accommodate the large structure produced
by the PRD and $J$-state interference effects and the fine structure in the cores of the lines \citep[see][for the details]{belluzzi_trujillo_mg12}. 
Such a large flexibility facilitates that the fits can be done with a very sparse $\mathbf{w}$ vector. The
results are shown in Fig. \ref{fig:mgii_belluzzi} for different noise levels parameterized
with the standard deviation of the Gaussian noise indicated in the lower right corner of
each panel. The number of basis functions for each case and their associated evidence
ratio with respect to the no-signal model is shown in the upper part of each panel. The results
indicate that the signal is nicely recovered with our method and that it is strongly in favor
of the presence of signal (relatively large evidence ratios) even for signal-to-noise (S/N) ratios on the range 1-3. The 
mean of the predictive distribution given by Eq. (\ref{eq:predictive}) shown with a red solid curve (and its associated
standard deviation, shown in dashed red curves) is a very good representation of the underlying synthetic signal.

\begin{figure*}[!t]
\includegraphics[width=0.5\textwidth,angle=-90]{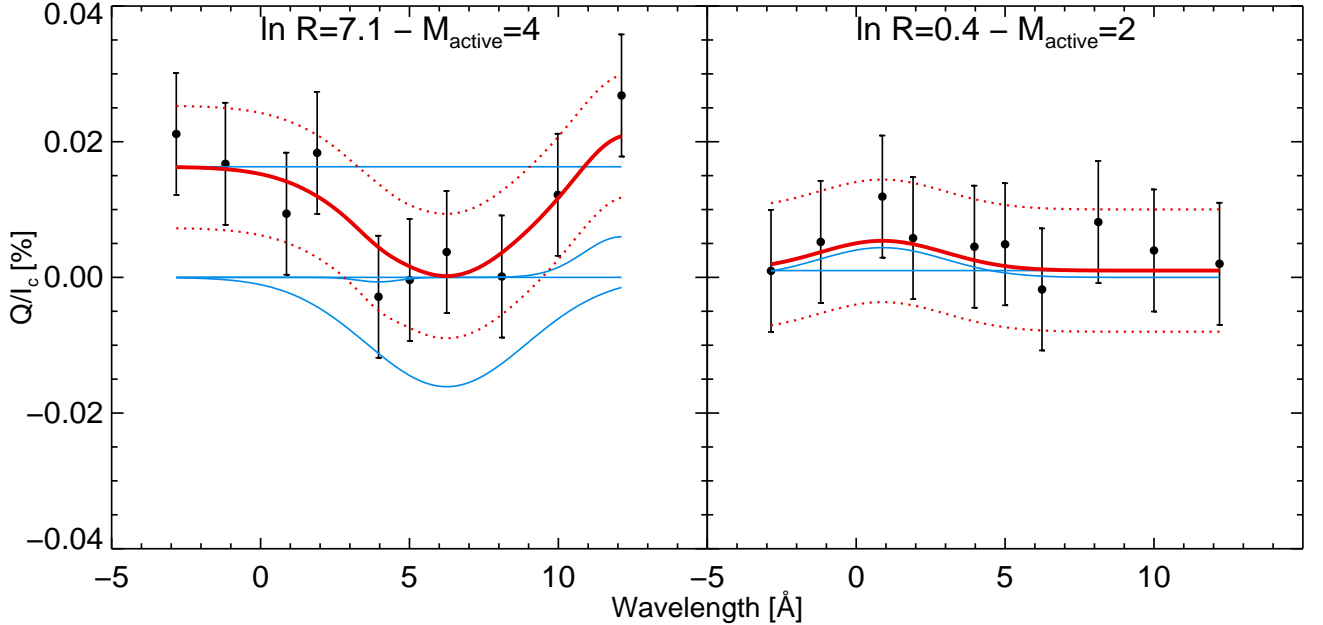}
\caption{Application of the signal detection scheme to the linear polarization signals
in the Mg \textsc{ii} h and k lines observed by \cite{henze87}. The dots display the
observations, with their associated Gaussian error bars. The solid red curve is the mean
of the predictive distribution, together with the error bars in red dotted lines. The blue curves display the
contribution of each individual kernel function.\label{fig:mgii}}
\end{figure*}

\subsubsection{Linear polarization in the He \textsc{i} 10830 \AA\ multiplet}
A second example showing the ability of our scheme to detect signals consists of
a synthetic linear polarization profile calculated with Hazel \citep{asensio_trujillo_hazel08}
for the 10830 \AA\ multiplet of neutral helium. The profiles are obtained at the solar
disk center with a magnetic field that is parallel to the surface with a strength
of 100 G. This is a typical configuration in which the Hanle effect generates
linear polarization in the forward scattering geometry due to a symmetry breaking
effect \citep[see][]{trujillo_nature02}. The
slab of He \textsc{i} atoms is assumed to be located at a height of $\sim$6000 km and
the optical depth measured in the red component of the multiplet (the one centered at $\sim$10830.5 \AA)
is 1.25. The width of the line is set to 8 km s$^{-1}$. Synthetic observations are generated
by adding different noises with standard deviations shown in the lower right corner of each 
panel (the error bar is also shown on the lower left corner). The basis set chosen for
the signal detection algorithm is made of Gaussian functions with widths between
0.3 and 1 \AA\ in 20 steps. Since the amplitude of the $Q/I_c$ signals (with $I_c$ being
the intensity at the continuum nearby) is $\sim$0.25\%
in the blue component and $\sim$0.5\% in the red component, the noises we have
considered are equivalent to S/N between 1 and 5 in the
red component and between 0.5 and 2.5 in the blue component. According to the results,
displayed in Fig. \ref{fig:hei_theory}, 
the non-parametric signal detection method gives an evidence ratio larger 
than 5 for the noisier case, strongly favoring the presence of a signal. The
mean of the predictive distribution (in red) is very similar to the synthetic one (in green) using
a very sparse solution with only 2 or 3 active basis functions.

\subsection{Real data}
\subsubsection{Linear polarization of the Mg \textsc{ii} h and k lines}
Given the difficulty of operating a spectropolarimeter on space, the only measurement of the linear 
polarization in the Mg \textsc{ii} h and k lines was carried out by \cite{henze87} using the Ultraviolet Spectrometer
and Polarimeter (UVSP) on the Solar Maximum Mission (SMM). The observations consisted of ten
wavelength samples across the h and k lines of Mg \textsc{ii} spanning a range of 15 \AA\ with
a slit length of 180$''$. They observed a region close to the solar limb and one at disk center. 
For symmetry reasons, the signal at disk center is expected to be zero (in the absence of a 
deterministic magnetic field in the resolution element), while it is expected to be
non-negligible close to the limb. Figure \ref{fig:mgii} shows, with dots, the observations
extracted from a scanned version of Fig. 1 in \cite{henze87}, in the left panel for the
observation at $\mu=0.15$ and
in the right panel for the observation at disk center. Each of the plotted points is
calculated as an average over all the observations of \cite{henze87} for a certain
wavelength bin. The error bar is estimated to be $\sigma=0.009$ which we consider fixed
and do not introduce it in the inference process (so $\sigma_\mathrm{MP}=0.009$). We apply the previous
formalism using a basis set composed of Gaussian functions centered at each observed point and with
widths ranging from 1 \AA\ to 5 \AA\ in 11 steps of 0.4 \AA\ plus a constant
function to allow for a continuum. Therefore, even though the number
of observations is $N=10$, the number of potentially active basis functions is $M=110$. 
Overfitting does not occur in our case because of the Bayesian treatment.
The solid red curve shows the mean of the predictive distribution while
the dashed red curves indicate its standard deviation (note that the predictive
distribution is Gaussian for each
predicted point). Computing the evidence ratio in the two cases, 
we find $\ln R=7.1$ for the profile close to the limb using only four active
kernels (shown as blue curves) and $\ln R=0.4$ at disk center using two active kernels (also shown
as blue curves). According to the standard
Jeffreys' scale, there is a really strong evidence for the presence of signal
in the observation close to the limb and inconclusive at disk center. Note also that the
solution is very sparse, using only $\sim$4\% of the potential basis functions for the
observation close to the limb and only $\sim$2\% for the observation at disk center.

\subsubsection{Linear polarization of the Ca \textsc{ii} H and K lines}
The second realistic example is the observation of the linear polarization signals of the H and K lines of Ca \textsc{ii}
in the UV. These signals have been acquired by \cite{gandorfer_atlas2_02} at an heliocentric
angle of $\mu=0.1$ and display an enormous amount of spectral signals that
are overlapped with the large-scale structure of the linear polarization of the two Ca \textsc{ii}
lines produced by superinterferences. We have resampled the profile at a spectral resolution of
$\sim$2 \AA\ to mimick a very low spectral resolution spectropolarimeter. The aim is to show that
it is possible to detect the linear polarization signal even at such low spectral resolutions
under the presence of large noise contaminations.

\begin{figure*}[!t]
\centering
\includegraphics[width=\textwidth]{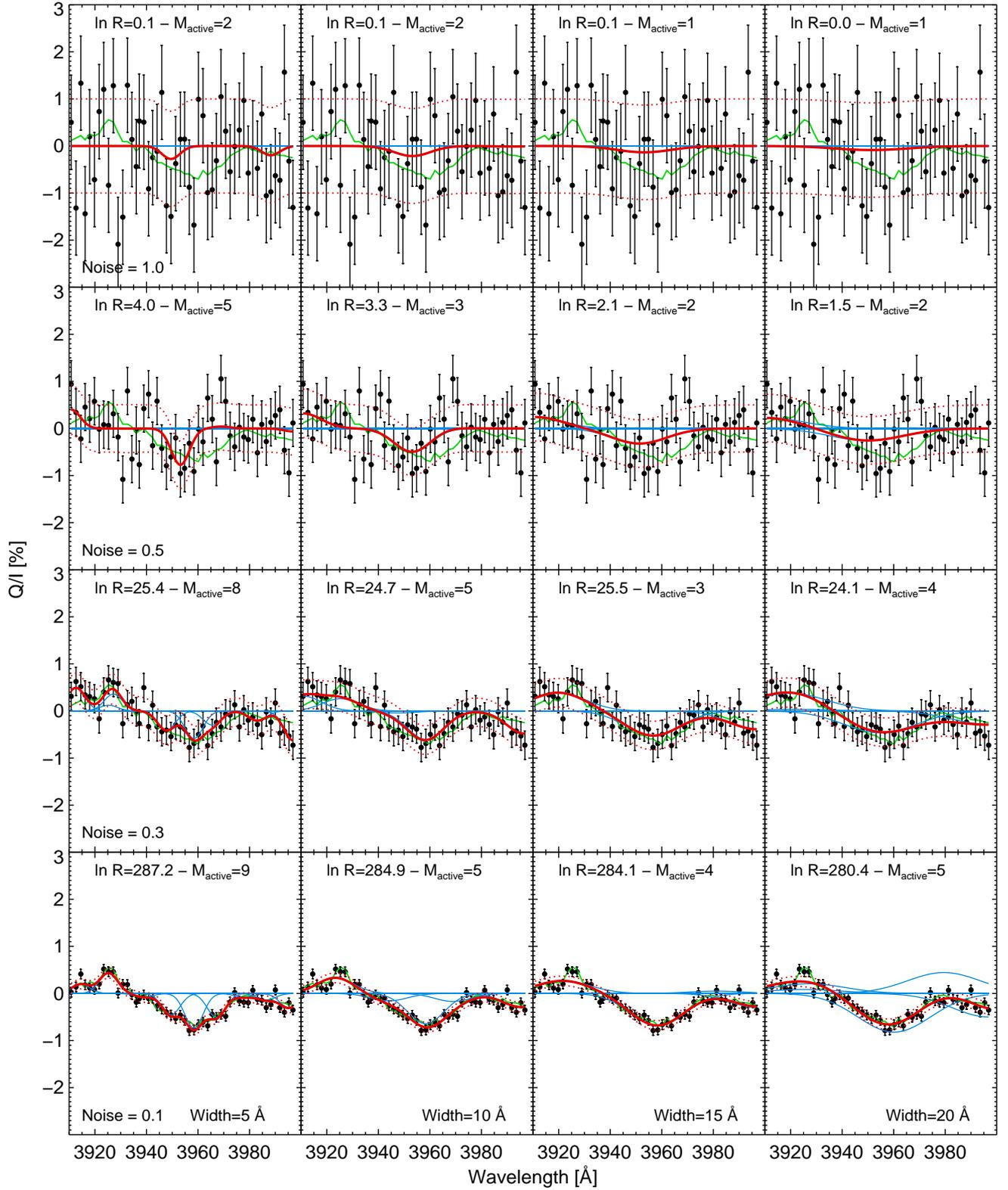}
\caption{Application to the linear polarization signals
in the Ca \textsc{ii} H and K lines observed in the atlas of \citep{gandorfer_atlas2_02} resampled
at 50 wavelength points and with different amounts of noise added for each row. The dots display the
observations, with their associated Gaussian error bars (with their standard deviation
indicated in the panels). Each column shows the results of the line detection using Gaussian functions
of different widths as basis functions. The solid red curve is the mean of the predictive
distribution, together with the range inside one standard deviation shown in red dotted lines. The blue curves display the
contribution of each individual kernel function. Each panel also displays the evidence ratio
and the number of active basis functions.\label{fig:caii_singlebasis}}
\end{figure*}

\begin{figure*}[!t]
\centering
\includegraphics[width=0.65\textwidth]{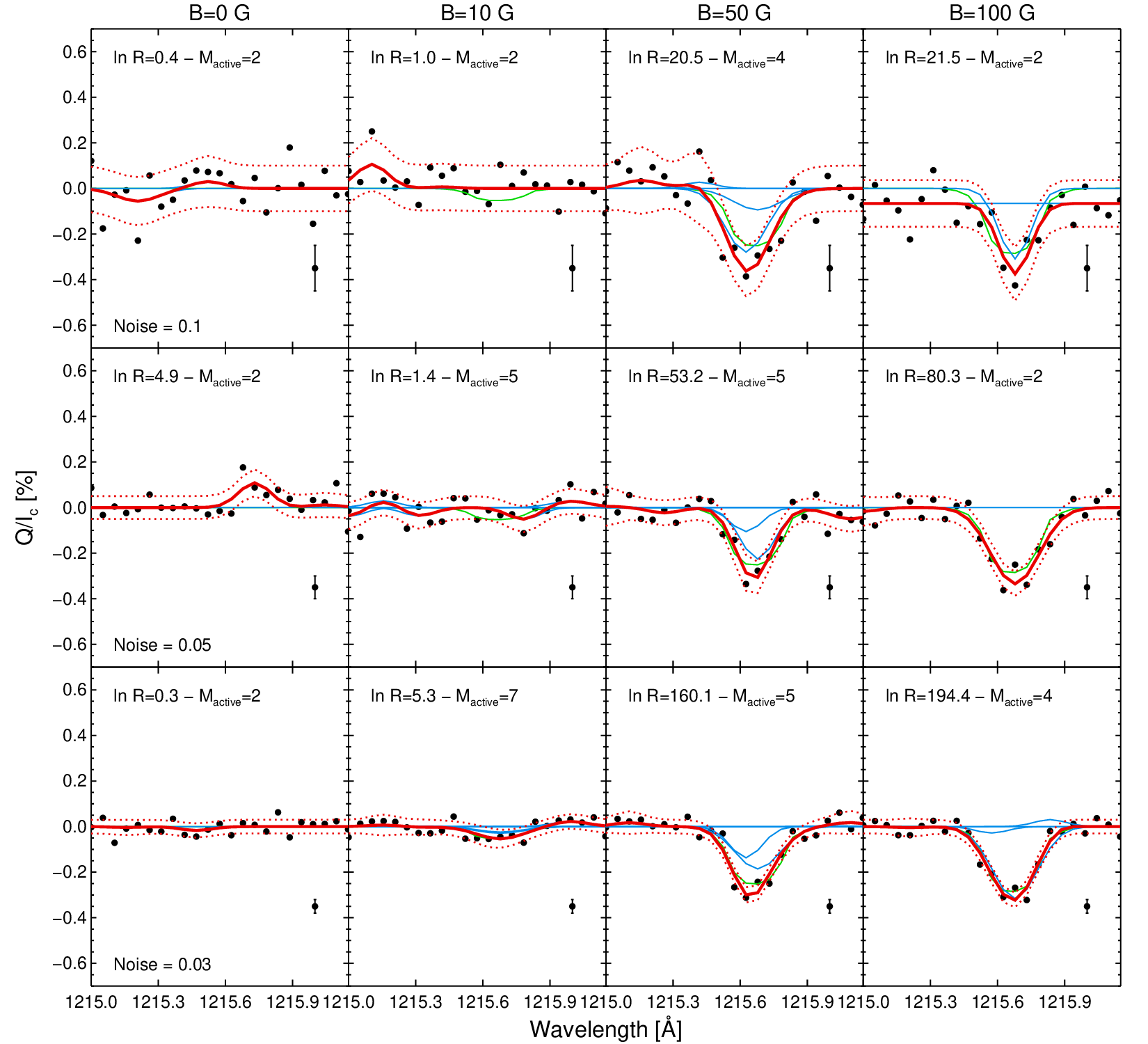}\\
\includegraphics[width=0.65\textwidth]{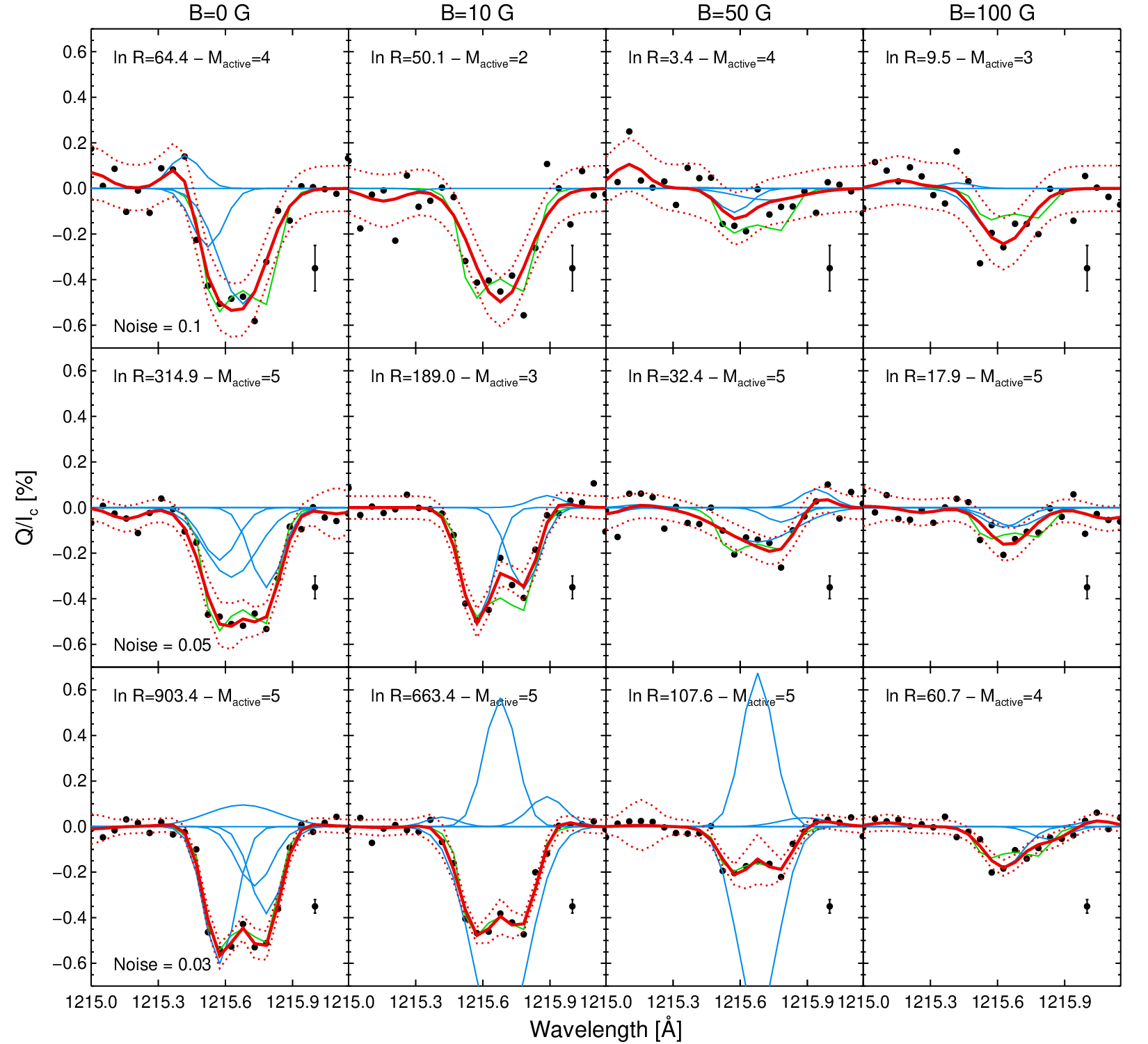}
\caption{Application of the signal detection scheme to the line-core linear polarization signals
estimated for the CLASP rocket experiment by \cite{trujillo_lya11}. The dots display the
observations, with their associated Gaussian error bars. The solid red curve is the mean
of the predictive distribution, together with the error bars in red dotted lines. The blue curves display the
contribution of each individual kernel function.\label{fig:clasp}}
\end{figure*}

We have carried out the signal detection procedure for four different levels
of Gaussian noise with different standard deviations, as shown in each
row of Fig. \ref{fig:caii_singlebasis}.
Given the original (resampled to low resolution) signals (shown in green in the figure), we 
contaminate them with Gaussian noise so that the S/N in the amplitude 
peaks of $Q/I$ range from 1 to 10, approximately. The signal detection is done with
basis sets composed of Gaussian functions of different widths (each column).
The results shown in Fig. \ref{fig:caii_singlebasis}
look very promising because, even for S/N as low as 1, we can reliably recover the original
signal, even though the observed signal is almost unrecognizable. The mean of the 
predictive distribution is surprisingly similar to a smoothed version of the green curve, specially when the basis width is large,
while many of the minute details of the signal can be estimated correctly if the noise is not too
large and the width of the Gaussian basis is small.

Concerning the evidence ratio, we find evidence for signal in all the cases. However, the
signal detection algorithm points to a moderate evidence for signal for the case with S/N$=1$.
The number of active Gaussian functions is usually smaller when the width is larger, with
an upper limit of 10 for the smallest considered noise level and width. In any case, we find that
the exact green curve is systematically inside one standard deviation of the predictive distribution.

\subsubsection{Linear polarization of the Ly$\alpha$ line with CLASP}
With the aim of investigating the magnetism of the upper chromosphere and transition
region of the Sun, the Chromospheric Lyman-Alpha Spectro-Polarimeter \cite[CLASP][]{kobayashi_clasp12} is a sounding rocket
proposed to carry out the first measurement of the linear polarization produced by scattering 
processes in the Ly$\alpha$ ultraviolet resonance line. A recent investigation
\cite{trujillo_lya11} indicates that the Ly$\alpha$ line should show measurable line-core
linear polarization either when observed at disk center or close to the solar limb. Additionally,
the linear polarization signal is sensitive to the magnetic ﬁeld strengths that are expected in the upper chromosphere
and transition region.

Because CLASP is mounted on a rocket, the total integration time is quite reduced. Consequently, the
final expected standard deviation of the noise (when taking into account the whole duration
of the mission of $\sim$5 min) is expected to be of the order of 0.03\% in units of
the monochromatic emission intensity of the line \cite[see][]{kobayashi_clasp12}. In order to
test the possibility of reliably detecting linear polarization signals with CLASP, we have
carried out the following experiment. We have used the $Q/I$ profiles computed by \cite{trujillo_lya11}
under the assumption of complete redistribution in frequency at two different positions in 
the solar disk and for four values of the strength of a horizontal magnetic field. The synthetic
curves are shown as green curves in Fig. \ref{fig:clasp}. The upper panel corresponds to an
observation at disk center, while the lower panel corresponds to an observation at $\mu=$0.3. The
observations have been corrupted with Gaussian noise of several standard deviation, from 0.03\% (the best
expected observation) up to 0.1\%. The signal detection is done with basis sets composed 
of Gaussian functions of widths between 0.1 and 0.3 \AA\ in steps of 0.01 \AA.

Given that the amplitude of the $Q/I$ signal depends on the magnetic
field strength, it is possible to find large and small evidence ratios for a fixed noise
variance. This is the case of the first row of the lower panel. The signal is clearly detected (large
value of the evidence ratio) up to fields below 50 G, but the case for 100 G gives no clear detection.
In fact, the specific value of the evidence ratio might change for different noise realizations.
When the standard deviation of the noise decreases, the method finds the signal in all
the cases with a very reduced set of basis functions. The predicted signal, shown as
a red curve, closely follows the synthetic one even in the cases with a reduced S/N.
Concerning the results at disk center, it is interesting to focus on the non-magnetic case. Given the symmetry of
the problem, the synthetic signal is strictly equal to zero. Our evidence ratios give no special preference
for the presence of a signal. From these results it seems that, if the $Q/I$ signal emerging from the solar atmosphere is
similar to the computed one, it is possible to detect it relaxing the CLASP requirements.

It is clear that the ultimate objective when detecting and extracting a signal from spectropolarimetric
observations is to infer the thermodynamic and magnetic properties of the plasma. To this end, the mean of the predictive distribution
can be used as a statistically meaningful estimation of the signal. Together with the mean, one has 
to add the error bars obtained from the standard deviation of the predictive
distribution. The main difficulty at this stage is to propose a suitable model for the polarimetric
signal from which one infers the thermal and magnetic properties. This is exactly the reason why we
have pursued a non-parametric scheme when we do not have a proper model for the expected signal
of interest. A straightforward way to proceed is to fit a suitable parametric model to the extracted 
signal with any standard least-squares algorithm. Given that this mixture of Bayesian and non-Bayesian
approaches surely does not make much sense, we are also in the process of studying a semiparametric 
(combination of parametric and non-parametric regressor) scheme that might give good results.

\section{Conclusions}
We have shown how a non-parametric Bayesian regression method 
can be applied to the problem of detecting a spectroscopic and/or spectropolarimetric signal that
is buried into the noise. The method is specially suited for analyzing signals whose spectral shape is
not known in advance. The output of the method is the evidence ratio between the model that
assumes a non-zero spectral signal and that assuming no signal is present. Without any additional computational
cost, the method also gives the predictive distribution, from where one can extract the
most probable regression and the corresponding error bars. This technique is appropriate for relaxing 
the noise requirements of observations where the shape of the signal is not known in
advance.

Our experiments in different spectral regions demonstrate that a signal corrupted with
Gaussian noise whose S/N of the order of 1 (or even smaller in some cases) can be efficiently detected and extracted using 
the non-parametric RVM method. We propose that a signal is detected when $\log R \gtrsim 2.5$ which,
according to the scale of \cite{jeffreys61}, corresponds to a moderate evidence in favor of the
presence of signal. Once the signal has been detected, signal extraction is carried out
by examining the mean of the predictive distribution and its associated standard deviation. The quality
of the signal extraction is obviously better when the signal is less buried into the noise. Summarizing,
we think that S/N$=1$ can be considered to be the lower limit for a reliable signal detection and
extraction.

Finally, we propose that this technique could be applied to the detection of the
{\em ultimate} property of light, its orbital angular momentum, from
astrophysical objects \citep{harwit03}, whose detection, if present, is going to
be challenging \citep{uribe11}.

\begin{acknowledgements}
We thank L. Belluzzi, J. \v{S}t\v{e}p\'an and J. Trujillo Bueno
for providing the synthetic profiles used in Figs. \ref{fig:mgii_belluzzi} and \ref{fig:clasp}.
Financial support by the Spanish Ministry of Economy and Competitiveness through projects AYA2010-18029 (Solar Magnetism and Astrophysical 
Spectropolarimetry) and Consolider-Ingenio 2010 CSD2009-00038 is gratefully acknowledged. AAR also acknowledges financial support through the Ram\'on y
Cajal fellowship. This research has 
benefited from discussions that were held at the
International Space Science Institute (ISSI) in Bern (Switzerland) in February 2010
as part of the International Working group \textit{Extracting information from spectropolarimetric
observations: comparison of inversion codes}.
\end{acknowledgements}

% \bibliographystyle{aa}
% \bibliography{/scratch/Dropbox/biblio}

\end{document}